\documentclass[hyper]{JHEP} 
\newcommand\fverb{\setbox\pippobox=\hbox\bgroup\verb}
\newcommand\fverbdo{\egroup\medskip\noindent%
            \fbox{\unhbox\pippobox}\ }

\newcommand\fverbit{\egroup\item[\fbox{\unhbox\pippobox}]}
\newbox\pippobox
\title{Semiclassical Strings on Curved Branes}
\author{Sagar Biswas\\
Department of Physics and Meteorology, \\
Indian Institute of Technology Kharagpur,\\
Kharagpur-721 302, INDIA \\
\email{biswas.sagar09iitkgp@gmail.com}}
\author{Kamal L Panigrahi\\
Department of Physics \& Meteorology,\\
Indian Institute of Technology Kharagpur,\\
Kharagpur-721302, INDIA,\\
and\\
The Abdus Salam International Centre for Theoretical Physics,\\
Strada Costiera 11, Trieste, ITALY\\
\email{panigrahi@phy.iitkgp.ernet.in}}
\abstract{We study semiclassical strings in the near horizon
geometry of certain curved branes. We investigate the rigidly
rotating strings in the near horizon geometry of NS5-branes
wrapped on $AdS_3 \times S^3$ and in the presence of background
NS-NS flux.  We study several string solutions corresponding to
giant magnon, single spike and more general folded strings for the
fundamental string in this background. We comment that in the
S-dual background the situation changes drastically.}

\begin{document}
\section{Introduction}
According to AdS/CFT duality
\cite{Maldacena:1997re},\cite{Gubser:1998bc},\cite{Witten:1998qj}
quantum closed string states in AdS should be dual to quantum
Super Yang-Mills (SYM) states on the boundary. More precisely,
this duality implies the equality between the AdS energy $E$ of
quantum closed string states (as function of effective string
tension $T$ and other quantum numbers like the angular momenta
$J_i$ on the sphere) and the dimension $\Delta$ of the
corresponding local SYM operators. Though the state-operator
matching is extremely difficult, but has been tractable in certain
limits, such as the large angular momentum limit, on both sides of
the duality \cite{Berenstein:2002jq},\cite{Gubser:2002tv}.
Further, it was observed that ${\cal N} =4$ SYM in planar limit
can be described by an integrable spin chain model where the
anomalous dimension of the gauge invariant operators were found in
\cite{Beisert:2005tm}, \cite{Minahan:2002ve},
\cite{Beisert:2003xu}, \cite{Beisert:2003yb},
\cite{Kazakov:2004qf}, \cite{Beisert:2004hm},
\cite{Arutyunov:2004vx}. In the dual picture, it was noticed that
the string theory is integrable in the semiclassical limit as
well, see for example \cite{Pohlmeyer:1975nb},
\cite{Minahan:2002rc}, \cite{Tseytlin:2004xa},
\cite{Hayashi:2007bq}, \cite{Okamura:2008jm}, hence providing
further insight into the AdS/CFT duality. However apart from few
`solvable' examples of AdS/CFT, in many cases the exact nature of
the boundary operators is not known, and hence it is interesting
to make calculations in the gravity side and then look for
possible operators on the boundary by invoking the duality map.
The study of rigidly rotating strings in semiclassical
approximation in the gravity side has been one of the interesting
areas of research in the last few years. In this connection a
large number of rotating and pulsating string solutions have been
studied in $AdS_5 \times S^5$, $AdS_4 \times CP^3$, orbifolded and
in the near horizon geometry of certain nonlocal string theory
backgrounds, see for example, \cite{Kruczenski:2004wg},
\cite{Bobev:2005cz}, \cite{Ryang:2004tq}, \cite{Dimov:2004xi},
\cite{Smedback:1998yn}, \cite{Hofman:2006xt}, \cite{Dorey:2006dq},
\cite{Chen:2006gea}, \cite{Bobev:2006fg},
\cite{Kruczenski:2006pk}, \cite{Chen:2006gq},
\cite{Hirano:2006ti}, \cite{Ryang:2006yq},
\cite{Maldacena:2006rv}, \cite{Kluson:2007qu},
\cite{Ishizeki:2007we}, \cite{Bobev:2007bm}, \cite{Hofman:2007xp},
\cite{Kluson:2007fr}, \cite{Ishizeki:2007kh}, \cite{Dorey:2007an},
\cite{Kruczenski:2008bs}, \cite{Lee:2008sk}, \cite{David:2008yk},
\cite{Kluson:2008gf}, \cite{Lee:2008ui}, \cite{Chen:2008qq},
\cite{Chen:2008vc}, \cite{Jain:2008mt}, \cite{Biswas:2011wu},
\cite{Panigrahi:2011be}, \cite{Biswas:2012wu}. However, more
recently, the integrability of the classical string motion in
curved $p$-brane background has been explored in
\cite{Stepanchuk:2012xi} in an attempt to see whether the full
string equations of motion is integrable. It is shown that moving
away from the throat geometry or in other words switching on the
brane charges actually destroyed the string integrability. Though
the point like string equations are in complete agreement with the
integrability, the equations describing an extended string in the
complete D-brane background, the integrable structure is lost. It
is definitely interesting to look for string equations of motion
in connection with integrability in various other situations.

Further to understand AdS/CFT like dualities in more general
backgrounds, arising out of near horizon geometries of various
branes in supergravities it is also interesting to look for
classical string equations of motion in the gravity side and make
statements about the integrability. It might help us in making
some observations about the dual theory which is apriori less
understood. Branes solutions with curved worldvolumes  have
widespread applications in string theory and black holes. In the
past they have been used to identify the non-perturbative states
of strings in lower dimensions in various string
compactifications. The curved brane solutions can be constructed
from the elementary solutions of the NS-NS sector which are
associated with conformally invariant sigma model
\cite{Giveon:1991jj},\cite{Russo:1995tj}. Indeed a large class of
solutions have been constructed in \cite{Papadopoulos:1999tw} by
using various dualities in string theory. We would like to study
semiclassical strings in some of these backgrounds. Specifically
we wish to study rigidly rotating strings in the near horizon
geometry of stack of NS5-brane with $AdS_3 \times S^3$
worldvolume. We study the most general form of the string equation
of motion and solve for the giant magnon and spiky like strings.
We further study a few general pulsating strings in this
background. Finally, we make some comments regarding F-string in
the S-dual background, namely the nature of the solutions to the
F-string equations of motion on a D5-brane wrapped on $AdS_3
\times S^3$. We remark that the possible non appearance of the
giant magnon or spike like solution is perhaps due to the non
integrability of the classical string equations of motion in the
D5-brane background.

The rest of the paper is organized as follows. In section-2, we
study rigidly rotating strings on NS5-brane wrapped on $AdS_3
\times S^3$ space. We find two limiting cases corresponding to
giant magnon and single spike solutions for the string in this
background. We present the regularized dispersion relations among
various conserved charges corresponding to the string motion.
Section-3 is devoted to the study of pulsating strings in this
background. In section 4, we make some remarks about the string
motion in the D5-brane wrapped on $AdS_3 \times S^3$ and conclude.
\section{Rotating String on Curved NS5-branes}  We
start with the solutions presented in \cite{Papadopoulos:1999tw}
that correspond to intersecting $NS1-NS1'-NS5-NS5'$ branes in
supergravity. The details of this background is given by the
following form of the metric, 2-form Neveu-Schwarz (NS) field
strength and dilaton \cite{Papadopoulos:1999tw}
\begin{eqnarray}
ds^2 &=& g_1^{-1}(x,y)(-dt^2+dz^2)+ H_5(x)dx^ndx^n +
H^{\prime}_5(y)dy^mdy^m, \\ \nonumber dB &=& dg_1^{-1}\wedge
dt\wedge dz + \star dH_5 + \star dH^{\prime}_5, \\ \nonumber
e^{2\phi} &=& \frac{H_5(x)H_5(y)}{g_1(x,y)} \ ,
\end{eqnarray}
where
\begin{equation}
[H^{\prime}_5(y)\partial^2_x + H_5(x)\partial^2_y]g_1(x,y)=0.
\end{equation}
A particular solution is given by
\begin{equation}
g_1(x,y)=H_1(x)H^{\prime}_1(y) \ ,
\end{equation}
with $H_{1,5} = 1 + \frac{Q_{1,5}}{x^2}$, and $H'_{1,5} = 1 +
\frac{Q'_{1,5}}{y^2}$, where $Q_{1,5}$ etc correspond to the
charges of F1 and NS5-brane respectively. The above solution
corresponds to the so called "dyonic string" generalization of the
$5_{NS}+ 5_{NS}$ solution in supergravity.

To continue further let us choose $H_1^{\prime}=1$, in which the
solution is just a direct product of $NS1+NS5$ configuration and
another $NS5-$ brane.
Further, in the near horizon limit as $x\rightarrow 0$, the metric
becomes,
\begin{equation}
ds^2=\frac{x^2}{Q_1}(-dt^2+dz^2) + Q_1\frac{dx^2}{x^2} +
Q_1d\Omega_3^2 + H_5^{\prime}(y)(dy^2+y^2d{\Omega_3^{\prime}}^2)
\end{equation}
where we have set for simplicity $H_1=H_5$. The resulting sigma
model describes a curved NS5-brane wrapped on $AdS_3 \times S^3$
and defines an exact CFT. We are interested in studying rigidly
rotating string on this NS5-brane in the near horizon geometry. In
the near horizon limit, $y\rightarrow0$, we get the following form
of the metric and NS-NS B-field.
\begin{equation}
ds^2 = Q'_5 (-\cosh^2\rho dt^2 + d\rho^2 + \sinh^2\rho d\varphi^2
+ d\Omega^2_3) + Q'_5(\frac{dy^2}{y^2}+ d{\Omega^{\prime}_3}^2) \
,
\end{equation}
with
$$d\Omega^2_3=d\theta^2_1 + \sin^2\theta_1 d\phi^2_1 + \cos^2\theta_1
d\psi_1^2 \ , $$ and
$$d{\Omega^{\prime}_3}^2=d\theta^2_2 + \sin^2\theta_2 d\phi^2_2 + \cos^2\theta_2
d\psi_2^2 .$$ The above metric is further supported by a NS-NS two
form field given by
$$B = 2Q'_5 \sin^2\theta_2 d\phi_2 \wedge d\psi_2 .$$ Note that for
convenience we have set $Q_1 = Q'_5$. This background is also
associated with an appropriate form of the dilaton whose explicit
form will not be needed here. To proceed further we make the
following change of variables $\chi=\ln y$. The final form of
metric and background field now takes the form
\begin{eqnarray}
&& ds^2 = Q'_5 (-\cosh^2\rho dt^2 + d\rho^2 + \sinh^2\rho
d\varphi^2 + d\theta^2_1 + \sin^2\theta_1 d\phi^2_1 +
\cos^2\theta_1 d\psi_1^2  \nonumber
\\ &&+ d\chi^2 + d\theta_2^2 + \sin^2\theta_2 d\phi_2^2 +\cos^2\theta_2
d\psi_2^2) \ , \>\>
B_{\phi_2\psi_2} = 2Q'_5 \sin^2\theta_2 . \nonumber \\
\end{eqnarray}
Note that we have used a different parameter to represent the
$AdS_3 \times S^3$ subspace.  We start by writing down the
Polyakov action of the F-string in the above background,
\begin{equation}
S=-\frac{\sqrt{\lambda}}{4\pi}\int d\sigma d\tau
[\sqrt{-\gamma}\gamma^{\alpha \beta}g_{MN}\partial_{\alpha} X^M
\partial_{\beta}X^N - e^{\alpha \beta}\partial_{\alpha} X^M
\partial_{\beta}X^N b_{MN}] \ ,
\end{equation}
where the 't Hooft coupling $\sqrt{\lambda} = Q'_5$,
$\gamma^{\alpha \beta}$ is the worldsheet metric and $e^{\alpha
\beta}$ is the antisymmetric tensor defined as $e^{\tau
\sigma}=-e^{\sigma \tau}=1$. Under conformal gauge (i.e.
$\sqrt{-\gamma}\gamma^{\alpha \beta}=\eta^{\alpha \beta}$) with
$\eta^{\tau \tau}=-1$, $\eta^{\sigma \sigma}=1$ and $\eta^{\tau
\sigma}=\eta^{\sigma \tau}=0$, the Polyakov action in the above
background takes the form,
\begin{eqnarray}
S &=& -\frac{\sqrt{\lambda}}{4\pi}\int d\sigma d\tau\Big[-\cosh^2
\rho({t^{\prime}}^2-\dot{t}^2) + {\rho^{\prime}}^2-\dot{\rho}^2 +
\sinh^2\rho({\varphi^{\prime}}^2-\dot{\varphi}^2) +
{\theta_1^{\prime}}^2-\dot{\theta_1}^2 \nonumber \\ &+&
\sin^2\theta_1({\phi_1^{\prime}}^2-\dot{\phi_1}^2)  +
\cos^2\theta_1({\psi_1^{\prime}}^2-\dot{\psi_1}^2) +
{\chi^{\prime}}^2-\dot{\chi}^2 +
{\theta_2^{\prime}}^2-\dot{\theta_2}^2 +
\sin^2\theta_2({\phi_2^{\prime}}^2-\dot{\phi_2}^2) \nonumber \\&+&
\cos^2\theta_2({\psi_2^{\prime}}^2-\dot{\psi_2}^2) -
4\sin^2\theta_2(\dot{\phi_2}\psi_2^{\prime}-\dot{\psi_2}\phi_2^{\prime})\Big]
\ ,
\end{eqnarray}
where `dots' and `primes' denote the derivative with respect to
$\tau$ and $\sigma$ respectively. For studying the rigidly
rotating strings we choose the following ansatz,
\begin{eqnarray}
\rho&=&\rho(y) ,\>\> t=\tau + h_1(y), \>\>\> \varphi=\mu(\tau +
h_2(y)),  \nonumber \\ \theta_1 &=& \theta_1(y), \>\>\>
\phi_1=\nu_1(\tau+g_1(y)), \>\>\> \psi_1=\omega_1(\tau+f_1(y)),
\nonumber \\ \theta_2 &=& \theta_2(y), \>\>\>
\phi_2=\nu_2(\tau+g_2(y)), \>\>\> \psi_2=\omega_2(\tau+f_2(y)),
\>\>\> \chi=\kappa\tau \ . \label{ansatz}
\end{eqnarray}
where $y=\sigma-v\tau$. Variation of the action with respect to
$X^M$ gives us the following equation of motion
\begin{eqnarray}
2\partial_{\alpha}(\eta^{\alpha \beta} \partial_{\beta}X^Ng_{KN})
&-& \eta^{\alpha \beta} \partial_{\alpha} X^M \partial_{\beta}
X^N\partial_K g_{MN} - 2\partial_{\alpha}(e^{\alpha \beta}
\partial_{\beta}X^N b_{KN}) \nonumber \\ &+& e^{\alpha \beta}
\partial_{\alpha} X^M \partial_{\beta} X^N\partial_K b_{MN}=0 \ ,
\end{eqnarray}
and variation with respect to the metric gives the two Virasoro
constraints,
\begin{eqnarray}
g_{MN}(\partial_{\tau}X^M \partial_{\tau}X^N +
\partial_{\sigma}X^M \partial_{\sigma}X^N)&=&0 \ , \nonumber \\
g_{MN}(\partial_{\tau}X^M \partial_{\sigma}X^N)&=&0 \ .
\end{eqnarray}
Next we have to solve these equations by the ansatz we have
proposed above in eqn. (\ref{ansatz}). Solving for $t, \varphi$ we
get,
\begin{eqnarray}
\frac{\partial h_1}{\partial y} =
\frac{1}{1-v^2}[\frac{c_1}{\cosh^2\rho}-v] ,\>\>  \frac{\partial
h_2}{\partial y} = \frac{1}{1-v^2}[\frac{c_2}{\sinh^2\rho}-v] \ .
\end{eqnarray}
Substituting these for $\rho$ equation we get,
\begin{eqnarray}
(1-v^2)^2\frac{\partial^2\rho}{\partial y^2} &=& \sinh \rho \cosh
\rho
[(1-\frac{c_1^2}{\cosh^4\rho})-\mu^2(1-\frac{c_2^2}{\sinh^4\rho})]
\ ,\nonumber \\ (1-v^2)^2(\frac{\partial \rho}{\partial y})^2 &=&
(1-\mu^2)\sinh^2 \rho + \frac{c_1^2}{\cosh^2\rho} - \frac{\mu^2
c_2^2}{\sinh^2\rho} + c_3 \ ,
\end{eqnarray}
where $c_1$, $c_2$ and $c_3$ are integration constants as well.
Similarly solving for $\phi_1$ and $\psi_1$ equations we get,
\begin{eqnarray}
\frac{\partial g_1}{\partial y} =
\frac{1}{1-v^2}[\frac{c_4}{\sin^2\theta_1}-v] \ , \frac{\partial
f_1}{\partial y} = \frac{1}{1-v^2}[\frac{c_5}{\cos^2\theta_1}-v] .
\end{eqnarray}
Substituting these for $\theta_1$ equation we get,
\begin{eqnarray}
(1-v^2)^2\left(\frac{\partial \theta_1}{\partial y}\right)^2 =
(\omega_1^2-\nu_1^2)\sin^2 \theta_1 - \frac{\nu_1^2
c_4^2}{\sin^2\theta_1} - \frac{\omega_1^2 c_5^2}{\cos^2\theta_1} +
c_6 \ , \label{theta1}
\end{eqnarray}
where $c_4$, $c_5$ and $c_6$ are integration constants. Again
solving for $\phi_2$ and $\psi_2$ equations we get,
\begin{eqnarray}
\frac{\partial g_2}{\partial y} =
\frac{1}{1-v^2}[\frac{c_7}{\nu_2\sin^2\theta_2}-
\frac{2\omega_2}{\nu_2}-v] \ ,  \>\>\> \frac{\partial
f_2}{\partial y} =
\frac{1}{1-v^2}[\frac{c_8}{\omega_2\cos^2\theta_2}-\frac{2\nu_2}{\omega_2}-v]
\ .\nonumber
\end{eqnarray}
Substituting these in $\theta_2$ equation we get,
\begin{eqnarray}
(1-v^2)^2\left(\frac{\partial \theta_2}{\partial y}\right)^2 =
3(\nu_2^2-\omega_2^2)\sin^2 \theta_2 -
\frac{c_7^2}{\sin^2\theta_2} - \frac{ c_8^2}{\cos^2\theta_2} + c_9
\ , \label{theta2}
\end{eqnarray}
where $c_7$, $c_8$ and $c_9$ are integration constants as well.
Now the Virasoro constraint $ T_{\tau\sigma} = 0$ gives
\begin{eqnarray}
&&(1-v^2)^2\Big[ \left(\frac{\partial \rho}{\partial y}\right)^2 +
\left(\frac{\partial \theta_1}{\partial y}\right)^2 +
\left(\frac{\partial \theta_2}{\partial y}\right)^2\Big] \nonumber \\
&=& \cosh^2\rho - \mu^2 \sinh^2\rho - \nu_1^2\sin^2\theta_1 -
\omega_1^2\cos^2\theta_1 - \nu_2^2 - \omega_2^2 -
3\nu_2^2\cos^2\theta_2 - 3\omega_2^2\sin^2\theta_2 \nonumber \\
&+& \frac{c_1^2}{\cosh^2\rho} - \frac{\mu^2 c_2^2}{\sinh^2\rho} -
\frac{\nu_1^2 c_4^2}{\sin^2\theta_1} -
\frac{\omega_1^2c_5^2}{\cos^2\theta_1}  -
\frac{c_7^2}{\sin^2\theta_2} -
\frac{c_8^2}{\cos^2\theta_2} + 4(\omega_2c_7 + \nu_2 c_8)  \nonumber \\
&+& \frac{1+v^2}{v}(-c_1 + \mu^2c_2 + \nu_1^2c_4 + \omega_1^2c_5 +
\nu_2c_7 + \omega_2c_8 - 2\nu_2\omega_2) \label{Virasoro-1}
\end{eqnarray}
Further, the Virasoro constraint $T_{\tau \tau} + T_{\sigma
\sigma}= 0$ gives
\begin{eqnarray}
&&(1-v^2)^2\Big[\left(\frac{\partial \rho}{\partial y}\right)^2 +
\left(\frac{\partial \theta_1}{\partial y}\right)^2 +
\left(\frac{\partial \theta_2}{\partial y}\right)^2\Big] \nonumber \\
&=& \cosh^2\rho - \mu^2 \sinh^2\rho - \nu_1^2\sin^2\theta_1 -
\omega_1^2\cos^2\theta_1 - \nu_2^2 - \omega_2^2 -
3\nu_2^2\cos^2\theta_2 - 3\omega_2^2\sin^2\theta_2 \nonumber \\
&+& \frac{c_1^2}{\cosh^2\rho} - \frac{\mu^2 c_2^2}{\sinh^2\rho} -
\frac{\nu_1^2 c_4^2}{\sin^2\theta_1} -
\frac{\omega_1^2c_5^2}{\cos^2\theta_1}  -
\frac{c_7^2}{\sin^2\theta_2} -
\frac{c_8^2}{\cos^2\theta_2} + 4(\omega_2c_7 + \nu_2 c_8)  \nonumber \\
&+& \frac{4v}{1+v^2}(-c_1 + \mu^2c_2 + \nu_1^2c_4 + \omega_1^2c_5
+ \nu_2c_7 + \omega_2c_8 - 2\nu_2\omega_2) -
\frac{(1-v^2)^2\kappa^2}{1+v^2} \ .
\end{eqnarray}
Subtracting the above two equations we get the following relation
among various parameters,
\begin{equation}
 -c_1 + \mu^2c_2 + \nu_1^2c_4 + \omega_1^2c_5 + \nu_2c_7 + \omega_2c_8 - 2\nu_2\omega_2 + \kappa^2 v =0.
\end{equation}
In what follows we will look at the two limiting cases
corresponding to giant magnon and single spike solutions for the
string in the curved NS5-brane near horizon background.
\subsection{Limiting cases}
Recall, we have from (\ref{theta1})
$$\left(\frac{\partial \theta_1}{\partial y}\right)^2 =
\frac{1}{(1-v^2)^2}\left[(\omega_1^2-\nu_1^2)\sin^2 \theta_1 -
\frac{\nu_1^2 c_4^2}{\sin^2\theta_1} - \frac{\omega_1^2
c_5^2}{\cos^2\theta_1} + c_6\right] $$
$\frac{\partial \theta_1}{\partial y} \rightarrow 0$ as $\theta_1
\rightarrow \frac{\pi}{2}$ implies $c_5=0$ and
$c_6=\nu_1^2c_4^2+\nu_1^2-\omega_1^2$, substituting this in the
above equation we get,
\begin{equation}
\frac{\partial \theta_1}{\partial
y}=\frac{\sqrt{\nu_1^2-\omega_1^2}}{1-v^2} \cot \theta_1
\sqrt{\sin^2\theta_1 - \alpha_1^2} \ ,
\end{equation}
where $\alpha_1^2=\frac{\nu_1^2c_4^2}{\nu_1^2-\omega_1^2}$.
Further, we have from (\ref{theta2})
$$\left(\frac{\partial \theta_2}{\partial y}\right)^2 =
\frac{1}{(1-v^2)^2}\left[3(\nu_2^2-\omega_2^2)\sin^2 \theta_2 -
\frac{c_7^2}{\sin^2\theta_2} - \frac{ c_8^2}{\cos^2\theta_2} +
c_9\right] .$$
Similarly, $\frac{\partial \theta_2}{\partial y} \rightarrow 0$ as
$\theta_2 \rightarrow \frac{\pi}{2}$ implies $c_8=0$ and
$c_9=3(\omega_2^2-\nu_2^2) + c_7^2$. Substituting this in the above
equation we get
\begin{equation}
\frac{\partial \theta_2}{\partial
y}=\frac{\sqrt{3(\omega_2^2-\nu_2^2)}}{1-v^2} \cot \theta_2
\sqrt{\sin^2\theta_2 - \alpha_2^2} \ ,
\end{equation}
where $\alpha_2^2=\frac{c_7^2}{3(\omega_2^2-\nu_2^2)}$.
Substituting the values of $\frac{\partial \theta_1}{\partial y}$
and $\frac{\partial \theta_2}{\partial y}$ with $c_5=c_8=0$ in
first Virasoro constraint (\ref{Virasoro-1}) we get,
\begin{eqnarray}
(1-v^2)^2\left(\frac{\partial \rho}{\partial y}\right)^2
= 1 + (1-\mu^2)\sinh^2\rho + \frac{c_1^2}{\cosh^2\rho} -
\frac{\mu^2 c_2^2}{\sinh^2\rho} -\alpha_3^2 \ ,
\end{eqnarray}
where
\begin{eqnarray}
\alpha_3^2 = \nu_1^2(1+c_4^2) +
\frac{1}{\nu_2^2}\{c_1-\mu^2c_2-\nu_1^2c_4- \kappa^2v\}^2 -
\{\kappa^2(1+v^2) + \nonumber
\\ + \kappa^2v(1+\frac{4\omega_2}{\nu_2}) + \nu_2^2 -
2\nu_2\omega_2\}
\nonumber \\
\end{eqnarray}
In limit $\frac{\partial \rho}{\partial y} \rightarrow 0$ as $\rho
\rightarrow 0$ implies $c_2=0$ and $c_1^2=\alpha_3^2-1$. Hence
\begin{equation}
\frac{\partial \rho}{\partial y} =
\frac{\sqrt{1-\mu^2}}{1-v^2}\tanh\rho \sqrt{\cosh^2\rho +
\alpha_4^2} \ ,
\end{equation}
where $\alpha_4^2=\frac{1-\alpha_3^2}{1-\mu^2}$. Looking at the
symmetry of the background of the near horizon of NS5-branes, a
number of conserved charges can be constructed as follows
\begin{eqnarray}
E &=& -\int\frac{\partial \mathcal{L}}{\partial \dot{t}}d\sigma=
\frac{\sqrt{\lambda}}{2\pi}\frac{1}{1-v^2}\int(\cosh^2\rho - c_1v)
d\sigma \ , \nonumber \\ S &=& \int\frac{\partial
\mathcal{L}}{\partial \dot{\varphi}} d\sigma
=\frac{\sqrt{\lambda}}{2\pi}\frac{\mu}{1-v^2}\int\sinh^2\rho
d\sigma \ , \nonumber \\ J_1 &=& \int\frac{\partial
\mathcal{L}}{\partial \dot{\phi_1}} d\sigma
=\frac{\sqrt{\lambda}}{2\pi}\frac{\nu_1}{1-v^2}\int(\sin^2\theta_1
- c_4v) d\sigma \ , \nonumber \\ J_2 &=& \int\frac{\partial
\mathcal{L}}{\partial \dot{\psi_1}} d\sigma
=\frac{\sqrt{\lambda}}{2\pi}\frac{\omega_1}{1-v^2}\int\cos^2\theta_1
d\sigma \ , \nonumber \\ K_1 &=& \int\frac{\partial
\mathcal{L}}{\partial \dot{\phi_2}} d\sigma
=\frac{\sqrt{\lambda}}{2\pi}\frac{1}{1-v^2}\int(3\nu_2\cos^2\theta_2
- 3\nu_2 - c_7 v) d\sigma \ , \nonumber \\ K_2 &=&
\int\frac{\partial \mathcal{L}}{\partial \dot{\psi_2}} d\sigma
=\frac{\sqrt{\lambda}}{2\pi}\frac{1}{1-v^2}\int(4\omega_2 +
2\nu_2v - 2c_7 - 3\omega_2\cos^2\theta_2) d\sigma \ , \nonumber \\
P &=& \int\frac{\partial \mathcal{L}}{\partial \dot{\chi}} d\sigma
= \frac{\sqrt{\lambda}}{2\pi}\kappa \int d\sigma \ . \nonumber \\
\end{eqnarray}
Also we have the following relation among various integration
constants
\begin{equation}
c_7=\frac{1}{\nu_2}[c_1 - \nu_1^2c_4 - \kappa^2 v + 2 \nu_2
\omega_2]
\end{equation}
\subsection{Single spike}
Let us look at various solutions to the string equations of motion
derived in the last section with appropriate choice of integration
constant. First, we choose $c_1=c_4=v$. Now the conserved
quantities become,
\begin{eqnarray}
E &=& \frac{\sqrt{\lambda}}{2\pi}\frac{1}{1-v^2}\int(\cosh^2\rho -
v^2) d\sigma , \\ \nonumber S &=&
\frac{\sqrt{\lambda}}{2\pi}\frac{\mu}{1-v^2}\int\sinh^2\rho d\sigma , \\
\nonumber J_1 &=&
\frac{\sqrt{\lambda}}{2\pi}\frac{\nu_1}{1-v^2}\int(\sin^2\theta_1
- v^2) d\sigma , \\ \nonumber J_2 &=&
\frac{\sqrt{\lambda}}{2\pi}\frac{\omega_1}{1-v^2}\int\cos^2\theta_1
d\sigma ,
\\ \nonumber K_1 &=&
\frac{\sqrt{\lambda}}{2\pi}\frac{1}{1-v^2}\int(3\nu_2\cos^2\theta_2
- 3\nu_2 - c_7 v) d\sigma , \\ \nonumber K_2 &=&
\frac{\sqrt{\lambda}}{2\pi}\frac{1}{1-v^2}\int(4\omega_2 + 2\nu_2v
- 2c_7 - 3\omega_2\cos^2\theta_2) d\sigma \\ \nonumber P &=&
\frac{\sqrt{\lambda}}{2\pi}\kappa \int d\sigma \ .
\end{eqnarray}
Also the relation among the integration constants now becomes
\begin{equation}
c_7=\frac{1}{\nu_2}[v(1 - \nu_1^2 - \kappa^2) + 2 \nu_2
\omega_2] \ .
\end{equation}
It is straightforward to see that the among various conserved
charges we get the following relations,
\begin{equation}
E-\frac{S}{\mu} = \frac{J_1}{\nu_1} + \frac{J_2}{\omega_1}
\end{equation}
and
\begin{eqnarray}
\frac{K_1}{\nu_2} + \frac{K_2}{\omega_2} =  \frac{1}{1-v^2}
\frac{1}{\kappa}\left[\frac{\omega_2+2\nu_2v}{\omega_2} -
\frac{(\omega_2v + 2\nu_2)(2\nu_2\omega_2 +v(1-\nu_1^2-
\kappa^2)}{\nu_2^2\omega_2} \right] P  \ .\nonumber \\
\end{eqnarray}
To find the explicit relation among various conserved charges
which looks like the spiky string, we now write the explicit
expression of the conserved charges. Now
\begin{equation}
J_1=\frac{\sqrt{\lambda}}{\pi}
\frac{\nu_1}{\sqrt{\nu_1^2-\omega_1^2}}[(1-v^2)
\int_{\frac{\pi}{2}}^{\arcsin(\alpha_1)} \frac{\sin \theta_1
d\theta_1}{\cos \theta_1 \sqrt{\sin^2 \theta_1 - \alpha_1^2}}
-\int_{\frac{\pi}{2}}^{\arcsin(\alpha_1)} \frac{\sin \theta_1 \cos
\theta_1 d\theta_1}{ \sqrt{\sin^2 \theta_1 - \alpha_1^2}} ] \ .
\nonumber \\
\end{equation}
$J_1$ diverges, but on regularization we get,
\begin{equation}
(J_1)_{\rm{reg}}=\frac{\sqrt{\lambda}}{\pi}
\frac{\nu_1}{\sqrt{\nu_1^2-\omega_1^2}} \sqrt{1-\alpha_1^2} \ .
\end{equation}
On the other hand $J_2$ is finite and is written as
\begin{equation}
J_2=-\frac{\sqrt{\lambda}}{\pi}
\frac{\omega_1}{\sqrt{\nu_1^2-\omega_1^2}} \sqrt{1-\alpha_1^2} \ .
\end{equation}
Similarly, $K_1$ and $K_2$ both diverge, however the regularized
expressions are given by
\begin{equation}
(K_1)_{\rm{reg}}=-\frac{\sqrt{\lambda}}{\pi}
\frac{3\nu_2}{\sqrt{3(\omega_2^2-\nu_2^2)}} \sqrt{1-\alpha_2^2} \
,
\end{equation}
and
\begin{equation}
(K_2)_{\rm{reg}}=\frac{\sqrt{\lambda}}{\pi}
\frac{3\omega_2}{\sqrt{3(\omega_2^2-\nu_2^2)}} \sqrt{1-\alpha_2^2}
\ .
\end{equation}
Now the angle difference between the end points of the string is
given by
\begin{equation}
\Delta \phi_1 = \nu_1\int_{-\infty}^{\infty}dy\frac{\partial
g_1}{\partial y} = -2\arccos(\alpha_1) \ ,
\end{equation}
which implies $\alpha_1=\cos \frac{\Delta \phi_1}{2}$.
However, $\Delta \phi_2 =
\nu_2\int_{-\infty}^{\infty}dy\frac{\partial g_2}{\partial y}$
diverge, but the regularized expression is given by
\begin{equation}
(\Delta \phi_2)_{{\rm reg}}=-2\arccos(\alpha_2) \ ,
\end{equation}
which implies $\alpha_2 = \cos \frac{(\Delta \phi_2)_{{\rm
reg}}}{2}$. In terms of $\Delta \phi_1$ and $(\Delta \phi_2)_{{\rm
reg}}$ we can express,
\begin{eqnarray}
(J_1)_{{\rm reg}} = \frac{\sqrt{\lambda}}{\pi}
\frac{\nu_1}{\sqrt{\nu_1^2-\omega_1^2}} \sin \frac{\Delta
\phi_1}{2} \ , \>\>\>\>  J_2 = -\frac{\sqrt{\lambda}}{\pi}
\frac{\omega_1}{\sqrt{\nu_1^2-\omega_1^2}} \sin \frac{\Delta
\phi_1}{2} , \nonumber \\
\end{eqnarray}
and they satisfy the relation,
\begin{equation}
(J_1)_{{\rm reg}} = \sqrt{J_2^2 +
\left(\frac{\lambda}{\pi^2}\right) \sin^2 \frac{\Delta \phi_1}{2}}
\ . \label{spike1}
\end{equation}
This relation looks precisely like the single spike dispersion
relation with two spins on $R\times S^3$ \cite{Ishizeki:2007we}.
Now,
\begin{eqnarray}
(K_1)_{\rm{reg}} = -\frac{\sqrt{\lambda}}{\pi}
\frac{3\nu_2}{\sqrt{3(\omega_2^2-\nu_2^2)}} \sin \frac{(\Delta
\phi_2)_{{\rm reg}}}{2} \ , \>\>\>\>  (K_2)_{\rm{reg}} =
\frac{\sqrt{\lambda}}{\pi}
\frac{3\omega_2}{\sqrt{3(\omega_2^2-\nu_2^2)}} \sin \frac{(\Delta
\phi_2)_{{\rm reg}}}{2} \ ,\nonumber \\
\end{eqnarray}
and they satisfy the relation,
\begin{equation}
(K_2)_{\rm{reg}} = \sqrt{(K_1^2)_{\rm{reg}} +
3\left(\frac{\lambda}{\pi^2}\right) \sin^2 \frac{(\Delta
\phi_2)_{\rm{reg}}}{2}}  \ .\label{spike2}
\end{equation}
We wish to mention that due to the presence of the background
$B$-field in the metric which is essentially the volume form of
the three sphere in the transverse space, we get a factor of 3 in
the dispersion relation in (\ref{spike2}) as compared to
(\ref{spike1}). We also have energy $E$ and spin $S$ of AdS space
as conserved quantities, which are diverging. However the
regularized expressions are given by
\begin{equation}
E_{\rm{reg}} = {\left(\frac{S}{\mu}\right)}_{\rm{reg}} =
-\frac{\sqrt{\lambda}}{\pi}
\frac{\sqrt{1+\alpha_4^2}}{\sqrt{1-\mu^2}} \ .
\end{equation}
So they satisfy
\begin{equation}
E_{\rm{reg}} - {\left(\frac{S}{\mu}\right)}_{\rm{reg}}=0 \ .
\end{equation}
The regularized spin can be rewritten as,
\begin{equation}
\frac{S_{\rm{reg}}}{\mu} =
\sqrt{S^2_{\rm{reg}}+\frac{\lambda}{\pi^2}(1+\alpha_4^2)} \ .
\end{equation}
The time difference $\Delta t$ between the end point of the string
can be defined as,
\begin{eqnarray}
\Delta t &=& \int_{-\infty}^{\infty}\frac{\partial h_1}{\partial
y}dy \nonumber \\ &=&
-\frac{2v}{\sqrt{1-\mu^2}}[\int_0^{\infty} \frac{\sinh\rho d\rho}{\cosh\rho
\sqrt{\cosh^2\rho+\alpha_4^2}} ] \ ,\nonumber \\
\end{eqnarray}
which is finite and is given by,
\begin{equation}
(\Delta t) =
-\frac{2v}{\sqrt{\alpha_3^2-1}}\arcsin\left(\frac{\sqrt{\alpha_3^2-1}}{\sqrt{1-\mu^2}}\right)
\ ,
\end{equation}
which implies
$$\frac{\sqrt{\alpha_3^2-1}}{\sqrt{1-\mu^2}}=-\sin(\frac{\Delta t\sqrt{\alpha_3^2-1}}{2v}).$$
In terms of $\Delta t$ we can express,
\begin{equation}
E_{\rm{reg}}=\frac{S_{\rm{reg}}}{\mu} =
\sqrt{S^2_{\rm{reg}}+\left(\frac{\lambda}{\pi^2}\right) \cos^2
\left(\frac{\Delta t\sqrt{\alpha_3^2-1}}{2v}\right)} \ .
\end{equation}
\subsection{Magnon Case}
In this case, let us choose $c_1=c_4=\frac{1}{v}$. Then the
conserved quantities become,
\begin{eqnarray}
E &=& \frac{\sqrt{\lambda}}{2\pi}\frac{1}{1-v^2}\int \sinh^2\rho
d\sigma \ , \nonumber \\ S &=&
\frac{\sqrt{\lambda}}{2\pi}\frac{\mu}{1-v^2}\int\sinh^2\rho
d\sigma \ , \nonumber \\ J_1 &=&
-\frac{\sqrt{\lambda}}{2\pi}\frac{\nu_1}{1-v^2}\int \cos^2\theta_1
d\sigma \ , \nonumber \\ J_2 &=&
\frac{\sqrt{\lambda}}{2\pi}\frac{\omega_1}{1-v^2}\int\cos^2\theta_1
d\sigma \ , \nonumber \\ K_1 &=&
\frac{\sqrt{\lambda}}{2\pi}\frac{1}{1-v^2}\int(3\nu_2\cos^2\theta_2
- 3\nu_2 - c_7 v) d\sigma \ , \nonumber \\ K_2 &=&
\frac{\sqrt{\lambda}}{2\pi}\frac{1}{1-v^2}\int(4\omega_2 + 2\nu_2v
- 2c_7 - 3\omega_2\cos^2\theta_2) d\sigma \ , \nonumber \\ P &=&
\frac{\sqrt{\lambda}}{2\pi}\kappa \int d\sigma \ . \nonumber \\
\end{eqnarray}
Also the relation among the integration constants now become
\begin{equation}
c_7=\frac{1}{\nu_2}\left[\frac{1}{v}(1 - \nu_1^2) - \kappa^2 v + 2
\nu_2 \omega_2\right] \ .
\end{equation}
Among the conserved quantities we get the following relations,
\begin{equation}
E-\frac{S}{\mu} = \frac{J_1}{\nu_1} + \frac{J_2}{\omega_1}=0 \ ,
\end{equation}
and
\begin{equation}
\frac{K_1}{\nu_2} + \frac{K_2}{\omega_2} = \frac{1}{1-v^2}
\frac{1}{\kappa}\left[\frac{\omega_2+2\nu_2v}{\omega_2} -
\frac{(\omega_2v + 2\nu_2)(2\nu_2\omega_2 -
\kappa^2v + \frac{1}{v}(1-\nu_1^2))}{\nu_2^2\omega_2} \right] P \ .
\end{equation}
The explicit expression of spin $S$ associated with $AdS$ is
diverging, but the regularized form is,
\begin{equation}
\frac{S_{\rm{reg}}}{\mu} = -\frac{\sqrt{\lambda}}{\pi}
\frac{\sqrt{1+\alpha_4^2}}{\sqrt{1-\mu^2}}
\end{equation}
This can be rewritten as,
\begin{equation}
\frac{S_{\rm{reg}}}{\mu} =
\sqrt{S^2_{\rm{reg}}+\frac{\lambda}{\pi^2}(1+\alpha_4^2)}
\end{equation}
The time difference $\Delta t$ between the end point of the string
can be defined as,
\begin{eqnarray}
\Delta t &=& \int_{-\infty}^{\infty}\frac{\partial h_1}{\partial
y}dy \nonumber \\ &=&
\frac{2}{\sqrt{1-\mu^2}}[(\frac{1}{v}-v)\int_0^{\infty}
\frac{\cosh\rho d\rho}{\sinh\rho \sqrt{\cosh^2\rho+\alpha_4^2}} -
\frac{1}{v}\int_0^{\infty} \frac{\sinh\rho d\rho}{\cosh\rho
\sqrt{\cosh^2\rho+\alpha_4^2}} ] \ ,\nonumber \\
\end{eqnarray}
which diverges, however the regularized $(\Delta t)_{\rm{reg}}$ is
,
\begin{equation}
(\Delta t)_{\rm{reg}} =
-\frac{2}{v\sqrt{\alpha_3^2-1}}\arcsin\left(\frac{\sqrt{\alpha_3^2-1}}{\sqrt{1-\mu^2}}\right)
\ ,
\end{equation}
which implies
$$\frac{\sqrt{\alpha_3^2-1}}{\sqrt{1-\mu^2}}=-\sin(\frac{\Delta t_{\rm{reg}}v\sqrt{\alpha_3^2-1}}{2}).$$
In terms of $(\Delta t)_{{\rm reg}}$ we can express,
\begin{equation}
\frac{S_{\rm{reg}}}{\mu} =
\sqrt{S^2_{\rm{reg}}+\left(\frac{\lambda}{\pi^2}\right) \cos^2
(\frac{\Delta t_{\rm{reg}}v\sqrt{\alpha_3^2-1}}{2})} \ .
\end{equation}
Again the angle difference $\Delta \phi_1$ is defined as,
\begin{eqnarray}
&&\Delta \phi_1 = \nu_1 \int_{-\infty}^{\infty}\frac{\partial
g_1}{\partial y}dy \nonumber \\ &=&
\frac{2\nu_1}{\sqrt{\nu_1^2-\omega_1^2}}\left[\frac{1}{v}
\int^{\frac{\pi}{2}}_{\arcsin(\alpha_1)}\frac{\cos\theta_1
d\theta_1}{\sin\theta_1\sqrt{\sin^2\theta_1-\alpha_1^2}} +
\left(\frac{1}{v}-v\right)
\int^{\frac{\pi}{2}}_{\arcsin(\alpha_1)}\frac{\sin\theta_1
d\theta_1}{\cos\theta_1\sqrt{\sin^2\theta_1-\alpha_1^2}} \right],
\nonumber \\
\end{eqnarray}
diverges. After excluding the divergence part, we get the
regularized $\Delta \phi_1$,
\begin{equation}
(\Delta \phi_1)_{\rm{reg}} = -2\arcsin(\alpha_1) \ ,
\end{equation}
which implies $\alpha_1 = -\sin
\frac{(\Delta\phi_1)_{\rm{reg}}}{2}$. The angular momentum $J_2$
is given by,
\begin{equation}
J_2=-\frac{\sqrt{\lambda}}{\pi}
\frac{\omega_1}{\sqrt{\nu_1^2-\omega_1^2}} \cos
\frac{(\Delta\phi_1)_{reg}}{2} \ ,
\end{equation}
which can be rewritten as,
\begin{equation}
\frac{J_2 \nu_1}{\omega_1} = \sqrt{J_2^2 +
\frac{\lambda}{\pi^2}\cos^2 \frac{(\Delta\phi_1)_{reg}}{2} } =
\frac{(J_2)_{reg}}{\omega_1} \ .
\end{equation}
Therefore we can write the giant magnon dispersion relation as,
\begin{eqnarray}
(E-J_1)_{{\rm reg}} &=& \frac{S_{{\rm reg}}}{\mu} +
\frac{(J_2)_{{\rm reg}}}{\omega_1} \nonumber \\ &=&
\sqrt{S^2_{{\rm reg}}+\frac{\lambda}{\pi^2} \cos^2 (\frac{\Delta
t_{{\rm reg}}v\sqrt{\alpha_3^2-1}}{2})} + \sqrt{J_2^2 +
\frac{\lambda}{\pi^2}\cos^2 \frac{(\Delta\phi_1)_{{\rm reg}}}{2}}
\ .
\nonumber \\
\end{eqnarray}
One may like to find further relations among the other charges
such as $K_1$, $K_2$ and $\Delta \phi_2$ which are confined to the
transverse space of the NS-brane.
\section{Folded String} In this section we wish to study some
string solutions which are pulsating in the background of the near
horizon geometry of the curved NS5-branes and also contain some
extra angular momentum. To study folded strings on this background
we choose the following ansatz,
\begin{eqnarray}
\rho(\sigma)&=&\rho(\sigma+2\pi) ,\>\> t=\kappa \tau , \>\>\>
\varphi=\mu_1\tau , \>\>\> \chi=\mu_2\tau, \\ \nonumber
\theta_1(\sigma) &=& \theta_1(\sigma+2\pi), \>\>\>
\phi_1(\sigma)=\phi_1(\sigma+2\pi), \>\>\> \psi_1=\omega_1 \tau,
\\ \nonumber \theta_2(\sigma) &=& \theta_2(\sigma+2\pi), \>\>\>
\phi_2(\sigma)=\phi_2(\sigma+2\pi), \>\>\> \psi_2=\omega_2\tau \ .
\end{eqnarray}
The Polyakov action of the string, in the conformal gauge and with
these ansatz, becomes,
\begin{eqnarray}
S &=& -\frac{\sqrt{\lambda}}{4\pi}\int d\sigma d\tau\Big[\cosh^2
\rho \dot{t}^2 + {\rho^{\prime}}^2 - \sinh^2\rho \dot{\varphi}^2 +
{\theta_1^{\prime}}^2 + \sin^2\theta_1{\phi_1^{\prime}}^2  -
\cos^2\theta_1 \dot{\psi_1}^2  \nonumber \\  &-& \dot{\chi}^2
+ {\theta_2^{\prime}}^2 + \sin^2\theta_2{\phi_2^{\prime}}^2 -
\cos^2\theta_2 \dot{\psi_2}^2 + 4\sin^2\theta_2
\dot{\psi_2}\phi_2^{\prime}\Big] \ .
\end{eqnarray}
Solving for $\rho$, $\theta_1$ and $\theta_2$ equations we get,
\begin{eqnarray}
\rho^{\prime \prime} &=& \sinh\rho \cosh\rho (\kappa^2 - \mu_1^2)
\ , \nonumber \\ \theta_1^{\prime \prime} &=& \sin \theta_1 \cos
\theta_1 ({\phi_1^{\prime}}^2 + \omega_1^2) \ , \nonumber \\
\theta_2^{\prime \prime} &=& \sin \theta_2 \cos \theta_2
({\phi_2^{\prime}}^2 + \omega_2^2 + 4\omega_2 \phi_2^{\prime}) \ ,
\nonumber \\ \label{pulsating-1}
\end{eqnarray}
Again solving for $\phi_1$ and $\phi_2$ equations, we get,
\begin{eqnarray}
\frac{d}{d\sigma}(\phi_1^{\prime}\sin^2\theta_1) = 0 \ , \>\>\>
\frac{d}{d\sigma}(\phi_2^{\prime}\sin^2\theta_2) +
4\omega_2\sin\theta_2\cos\theta_2\frac{d\theta_2}{d\sigma} = 0 \ .
\end{eqnarray}
Integrating these two, we get
\begin{eqnarray}
\phi_1^{\prime} = \frac{c_1}{\sin^2\theta_1} \ , \>\>\>
\phi_2^{\prime} = \frac{c_2}{\sin^2\theta_2} -2\omega_2 \ ,
\end{eqnarray}
where $c_1$ and $c_2$ are integration constants. Substituting the
values of $\phi_1^{\prime}$ and $\phi_2^{\prime}$ in
(\ref{pulsating-1}) equations of motion and integrating them we
get,
\begin{eqnarray}
{\rho^{\prime}}^2 &=& (\kappa^2 - \mu_1^2) \sinh^2 \rho + c_3 \ ,
\nonumber \\ {\theta_1^{\prime}}^2 &=&
-\frac{c_1^2}{\sin^2\theta_1} + \omega_1^2\sin^2\theta_1 + c_4 \ ,
\nonumber \\ {\theta_2^{\prime}}^2 &=&
-\frac{c_2^2}{\sin^2\theta_2} -3 \omega_2^2\sin^2\theta_1 + c_5 \
, \nonumber \\
\end{eqnarray}
where $c_3$, $c_4$ and $c_5$ are integration constants as well.
Now from the Virasoro constraints, we get the following relation
among various integration constants,
\begin{equation}
-\kappa^2 - \mu_2^2 + \omega_1^2 - \omega^2_2 + 4 c_2 \omega_2 + c_3
+c_4 - c_5 = 0 \ .
\end{equation}
The conserved quantities in these case are given by,
\begin{eqnarray}
E &=& \frac{\sqrt{\lambda}\kappa}{2\pi} \int_0^{2\pi} d\sigma
\cosh^2\rho \ , \nonumber \\ S &=& \frac{\sqrt{\lambda}
\mu_1}{2\pi} \int_0^{2\pi} d\sigma \sinh^2\rho \,  \nonumber \\ J
&=& \frac{\sqrt{\lambda}\omega_1}{2\pi} \int_0^{2\pi} d\sigma
\cos^2\theta_1 \ , \nonumber \\ K &=& \frac{\sqrt{\lambda}}{2\pi}
\int_0^{2\pi} d\sigma (4\omega_2 - 3\omega_2 \cos^2\theta_2 -
2c_2) \ , \nonumber \\ P &=& \sqrt{\lambda} \mu_2 \ . \nonumber \\
\end{eqnarray}
We can choose $c_1=c_2=0$, so that we can express our result in
terms of elliptic functions as is the usual practice. In what
follows, we wish to study few subset of pulsating solutions.
\subsection{For $\theta_1 = \theta_2 =0$} In this section we wish
to studying the string which pulsates in the AdS$_3$ subspace and
which contains extra charges due to the transverse motion of the
string along the radial direction. We define the energy and spin
density as
\begin{eqnarray}
\mathcal{E} &=& \frac{E}{\sqrt{\lambda}} =
\frac{\kappa}{2\pi}\int_0^{2\pi}d\sigma \cosh^2\rho \ , \nonumber \\
\mathcal{S} &=& \frac{S}{\sqrt{\lambda}} =
\frac{\mu_1}{2\pi}\int_0^{2\pi}d\sigma \sinh^2\rho \ .
\end{eqnarray}
Hence they satisfy the relation
$$\frac{\mathcal{E}}{\kappa} - \frac{\mathcal{S}}{\mu_1} = 1$$
or,
\begin{equation}
\mathcal{E} = \kappa + \frac{\kappa}{\mu_1}\mathcal{S}
\end{equation}
Also we can define,
\begin{eqnarray}
\mathcal{J} = \frac{J}{\sqrt{\lambda}} = \omega_1 \ ,\>\>\>
\mathcal{K} = \frac{K}{\sqrt{\lambda}} = \omega_2 \ , \>\>\>
\mathcal{P} = \frac{P}{\sqrt{\lambda}} = \mu_2 \ .
\end{eqnarray}
Now we have,
\begin{eqnarray}
\rho^{\prime} &=& \frac{d\rho}{d\sigma} = \sqrt{c_3 + (\kappa^2 -
\mu_1^2)\sinh^2\rho}  \nonumber \\
\int_0^{2\pi} d\sigma &=& 4\int_0^{\rho_0} \frac{d\rho}{ \sqrt{c_3
+ (\kappa^2 - \mu_1^2)\sinh^2\rho}}
\end{eqnarray}
where $\rho_0$ corresponds to maximum value $\sinh \rho_0$.
Solving this we get,
\begin{equation}
\sqrt{\mu_1^2 - \kappa^2} = \frac{2}{\pi} K(q) \ , \label{spin-1}
\end{equation} where $K(q)$ is the elliptic function of first kind
with argument $q=\sqrt{\frac{c_3}{\kappa^2 - \mu_1^2}}$. Also we
have,
\begin{eqnarray}
\mathcal{E} =  \frac{\kappa}{2\pi}\int_0^{2\pi} d\sigma
\cosh^2\rho = \frac{4\kappa}{2\pi}\int_0^{\rho_0} \frac{d\rho
\cosh^2\rho}{ \sqrt{c_3 + (\kappa^2 - \mu_1^2)\sinh^2\rho}} \ .
\end{eqnarray}
Solving this integration we get,
\begin{equation}
\mathcal{E}=\frac{2\kappa}{\pi} \frac{E(q)}{\sqrt{\mu_1^2 -
\kappa^2}} \ ,\label{energy-1}
\end{equation}
where $E(q)$ is the elliptic function of second kind. Combining
the two equations (\ref{spin-1}) and (\ref{energy-1}) we get
\begin{eqnarray}
\kappa^2 &=& \left(\frac{K(q)}{E(q)}\mathcal{E}\right)^2 \ , \nonumber \\
\mu_1^2 &=& \left(\frac{K(q)}{E(q)}\mathcal{E}\right)^2 +
\frac{4}{\pi^2}(K(q))^2 \ .
\end{eqnarray}
Similar solutions were found in \cite{Beisert:2003ea}.
\subsection{For $\rho = \theta_2 =0$}
In this section we wish to study strings that pulsate in one of
the S$^3$ and at the same time have extra charges due to the
transverse motion of the string in the radial direction of the
NS5-brane. Note that the extra charges appear because of the
translational symmetry along the $\xi$ direction of the original
background. For this case we have,
\begin{eqnarray}
\mathcal{E} = \kappa \ , \mathcal{S} = 0 \ , \mathcal{K} =
\omega_2 \ ,\mathcal{P} = \mu_2 \ .
\end{eqnarray}
We also have,
\begin{equation}
\frac{d\theta_1}{d\sigma} = \sqrt{c_4 + \omega_1^2 \sin^2\theta_1}
\ ,
\end{equation}
which implies
\begin{equation}
\omega_1 = -\frac{2}{\pi}K(r) \ , \label{ang-mom-2}
\end{equation}
where $K(r)$ is the elliptic function of first kind with the argument $r=\frac{\sqrt{c_4}}{\omega_1}$
and
\begin{equation}
\mathcal{J} = \frac{\omega_1}{2\pi}\int_0^{2\pi} d\sigma
\cos^2\theta_1 \ ,
\end{equation}
which implies
\begin{equation}
\mathcal{J} = -\frac{2}{\pi} E(r)  \ , \label{energy-2}
\end{equation}
where $E(r)$ is the elliptic function of second kind. Combining
the two equations (\ref{ang-mom-2}) and (\ref{energy-2}) we get,
\begin{equation}
\omega_1^2 = \left(\frac{K(r)}{E(r)}\mathcal{J}\right)^2 \ .
\end{equation}
Similar solutions have been found in \cite{Beisert:2003ea}.
\subsection{For $\rho = \theta_1 = 0$} This is an interesting
case, where not only the string pulsates in one of the S$^3$, it
also has extra charges due to the transverse motion of the string
in the radial direction and furthermore there is a non-zero
B-which contributes the equations of motion. Hence the fundamental
string would know the presence of such field through the
energy-spin relationship. For this case we have,
\begin{eqnarray}
\mathcal{E} = \kappa \ , \mathcal{S} = 0 \ , \mathcal{J} =
\omega_1 \ , \mathcal{P} = \mu_2 \ .
\end{eqnarray}
We also have,
\begin{equation}
\frac{d\theta_2}{d\sigma} = \sqrt{c_5 - 3\omega_2^2
\sin^2\theta_2} \ ,
\end{equation}
which implies
\begin{equation}
\omega_1 = \frac{2}{\sqrt{3}\pi}K(s)
\end{equation}
where $K(s)$ is the elliptic function of first kind with the
argument $s=\frac{\sqrt{c_5}}{\sqrt{3}\omega_2}$ and
\begin{eqnarray}
\mathcal{K} &=& \frac{\omega_2}{2\pi}\int_0^{2\pi} d\sigma (4-3\cos^2\theta_2) \\ \nonumber &=& \frac{4\omega_2}{2\pi}
\Big[4\int_0^{\frac{\pi}{2}}\frac{d\theta_2}{\sqrt{c_5-3\omega_2^2\sin^2\theta_2}}
- 3\int_0^{\frac{\pi}{2}}\frac{d\theta_2
\cos^2\theta_2}{\sqrt{c_5-3\omega_2^2\sin^2\theta_2}} \Big] \ ,
\end{eqnarray}
which implies
\begin{equation}
\mathcal{K} = \frac{2}{\sqrt{3}\pi}[4K(s) -  3E(s)] \ ,
\end{equation}
where $E(s)$ is the elliptic function of second kind. Combining these two equations we can write,

$$\mathcal{K}=\left[4-3\frac{E(s)}{K(s)}\right]\omega_2 \ ,$$
or,
\begin{equation}
\omega_2^2 =
\frac{\mathcal{K}^2}{\left[4-3\frac{E(s)}{K(s)}\right]^2} \ .
\end{equation}
\section{Discussion and Conclusion} In this paper we have studied
semiclassical strings in the near horizon geometry of curved
NS5-branes, namely on the NS5-branes with $AdS_3 \times S^3$
worldvolume. We have found the most general solutions of the
equations of motion of the probe fundamental string in this
background and found out solutions corresponding to giant magnon,
single spike and furthermore the pulsating strings. We have found
out the dispersion relation among various conserved charges and
compare them with the existing ones. The novelty of these
solutions is that they contain the information about the
background NS-NS field. Further, the presence of the charge, $P$,
in the dispersion relation reflects the fact that the motion of
the string in the radial direction $\xi$ in the near horizon
geometry of NS5-branes is free. In the spirit of the
non-integrability of the classical strings in the generic
$p$-brane background, one can try to investigate the fundamental
string equations of motion in the S-dual background, i.e. the
D5-brane background wrapped on $AdS_3 \times S^3$, which is
presented in this paper. The details of the background is given by
\cite{Papadopoulos:1999tw}. By looking at the classical
integrability of the string solutions presented in this paper in
the NS5-brane background, one might be tempted to believe that
similar solution would appear because of the $AdS_3 \times S^3
\times S^3$ structure of the parent background (in the absence of
any brane charges). But we notice that while solving for the
equations of motion in the D5-brane background, it is not possible
to find simple or similar solution corresponding to the usual
giant magnon and single spike strings. This is perhaps a hint to
believe that F-string equations of motion are non-integrable in
the D5-brane background. However, we wish to remark that the
background solutions for the NS5-brane and D5-brane are similar
being related by S-duality, but this S-duality does not act on
classical string solutions in these backgrounds. Therefore,
classical string solutions would indeed be very different. Hence
in D5-brane background case there is no reason to expect
integrability of probe fundamental string equations. It would
perhaps be interesting to study D1-brane equations of motion in
the D5-brane background and look for exact solutions in the
context of integrability. We wish to come back to this issue in
future. \vskip .1in \noindent {\bf Acknowledgements:} We would
like to thank A. Tseytlin for some comments. KLP would like to
thank the Abdus Salam I.C.T.P, Trieste for hospitality under
Associate Scheme, where a part of this work was completed.

\end{document}